\documentclass[aps,twocolumn,showpacs]{revtex4}

\usepackage{amsmath}
\usepackage{graphicx}

\begin{document}
\title[]{Cold atom dynamics in a quantum optical lattice potential}

\author{Christoph Maschler}
\author{Helmut Ritsch}%
\affiliation{%
Institut f\"ur theoretische Physik, Universit\"at Innsbruck,
Technikerstr.~25, A-6020 Innsbruck,\\ Austria}%

\begin{abstract}
We study a generalized cold atom Bose Hubbard model, where the periodic optical potential is formed by a cavity field with quantum properties. On the one hand the common coupling of all atoms to the same mode introduces cavity mediated long range atom-atom interactions and on the other hand atomic backaction on the field introduces atom-field entanglement. This modifies the properties of the associated quantum phase transitions and allows for new correlated atom-field states including superposition of different atomic quantum phases. After deriving an approximative Hamiltonian including the new long range interaction terms we exhibit central physical phenomena at generic configurations of few atoms in few wells. We find strong modifications of population fluctuations and next-nearest neighbor correlations near the phase transition point.            

\end{abstract}


\maketitle

Laser fields can nowadays be routinely used to create tailored optical potentials for ultracold neutral atoms~\cite{lasercool}. Loading an atomic BEC into such a periodic standing light pattern allows to experimentally implement systems, which are well described by a Bose Hubbard Hamiltonian with externally controllable parameters~\cite{Fisher,Jaksch,Zwerger}. In some recent spectacular experiments the predicted Mott-insulator to superfluid quantum phase transition has been observed~\cite{Expmott}. As the light fields are normally intense and strongly detuned from any atomic transition, their properties can be safely approximated by prescribed classical fields independent of the atomic state. However, this is invalid if they are confined within an optical resonator. For a sufficient atom number $N$ and atom-field coupling $g$ the fields become dynamical quantities depending on the atoms. In addition in a high-$Q$ cavity the quantum properties of the field get important and the atoms move in quantized potentials~\cite{Domokos03,Griessner}. Ultimately this allows different states of the lattice field (e.g. different photon numbers) to be quantum correlated with different quantum phases of the atoms. As a striking example the atoms could be in a superposition of a Mott insulator and a superfluid state connected with a different cavity field amplitudes. Even in the classical field limit of high photon numbers all atoms see the same field state and thus we get new long range atom-atom interactions. Interestingly the idea of implementing such combination of cavity QED and a BEC has experimentally made such rapid progress recently, that its success can be expected very soon~\cite{BEC_Cav}. 

In this work we discuss the basic physical properties of such a generalized model of ultracold atoms in a periodic potential generated by a quantized field mode. In a first step we derive an approximate Hamiltonian analogous to the Bose Hubbard Hamiltonian including a quantized potential. Its basic physical implications are then exhibited in two cases: (a) a strongly damped cavity, where the field dynamics can be adiabatically eliminated, which leads to a rescaling of the coupling parameters and new long range atom-atom coupling terms and (b) a dynamical model where the field is approximated by its time dependent expectation value derived from a dynamical equation containing atomic expectation values.                  

\begin{figure}[htp]
\begin{center}
   \scalebox{0.30}[0.30]{\includegraphics{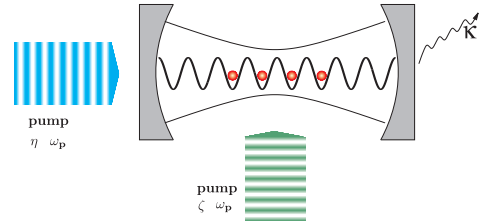}}
     \caption{\label{cavity}(color online). Sketch of setup}
    \nobreak\medskip
  \end{center}
\end{figure}

{\sl Model:} Let us consider $N$ two-level atoms with mass $m$ and transition frequency $\omega_a$ strongly interacting with a single standing wave cavity mode of frequency $\omega_c \ll \omega_a$. The system is coherently driven by a laser field with frequency $\omega_p \approx \omega_c$ through the cavity mirror with amplitude $\eta$. Alternatively the atoms are illuminated transversally to the cavity axes (see Fig.~\ref{cavity}) with amplitude $\zeta$. 

Including damping, the dynamics is given by the master equation for the atom field density operator
$ \dot{\varrho}=\frac{1}{i\hbar}[H,\varrho]+\mathcal{L}\varrho $,
where the Liouvillian $\mathcal{L}$ models dissipation. For large atom-pump detuning spontaneous emission is negligible and cavity loss $\kappa$ will be the dominant dissipative process, i.e., $\mathcal{L}\varrho=\kappa\left( 2a\varrho a^\dag - a^\dag a\varrho -\varrho a^\dag a\right) $. Here $a$ is the annihilation operator for a cavity photon. 

As convenient choice we rewrite the Hamiltonian in a second quantized form, where the direct interaction between the atoms is modeled by a pseudopotential and characterized by the s-wave scattering length $a_s$:

   \begin{eqnarray}\label{ham2ndquant} H&=&\int \mathrm{d}^3 x \Psi^\dag\left({\bf x}\right) H_0\Psi\left({\bf x}\right)\nonumber\\&+&\frac{1}{2}\frac{4\pi a_s\hbar^2}{m}\int\mathrm{d}^3 x \Psi^\dag\left({\bf x}\right)\Psi^\dag\left({\bf x}\right)\Psi
   \left({\bf x}\right)\Psi\left({\bf x}\right).\end{eqnarray}

Here $\Psi\left({\bf x}\right)$ is the field operator for the atoms and $H_0$ is the single-particle Hamiltonian in rotating-wave and dipole approximation:

  \begin{eqnarray} \nonumber
     H_0&=&\frac{p^2}{2m}-\hbar\Delta_a\sigma^+\sigma^--\hbar\Delta_c a^\dag a-i\hbar g(x)\left( \sigma^+a-\sigma^-a^\dag\right)\\&-&
     i\hbar h\left({\bf x}\right)\zeta\left( \sigma^+-\sigma^-\right)-i\hbar\eta\left( a-a^\dag\right), \label{ham}\end{eqnarray}

where $\Delta_a=\omega_p-\omega_a$ and $\Delta_c=\omega_p-\omega_c$ is the atom-pump and cavity-pump detuning, respectively. Along the cavity axis (x-direction) the atom-field coupling is set to $g(x)=g_0\cos (kx)$, while the transverse laser beam forms a standing wave with amplitude $h\left({\bf x}\right)=h_0\cos(k_py)$ in the y-direction, where we set $y=0$ for our 1D-considerations. In the regime of low saturation~\cite{masch04} (large $\Delta_a$) adiabatic elimination of the excited atomic state in~(\ref{ham}) then leads to:

  \begin{eqnarray}\nonumber H_0&=&\frac{p^2}{2m}+\cos^2(kx)\left(\hbar U_0 a^\dag a+V_{\textrm{cl}}\right)-\hbar\Delta_c a^\dag a\\&-&i\hbar\eta\left( a-a^\dag\right) +\hbar\eta_{
  \textrm{eff}}\cos(kx)\left( a+a^\dag\right)\label{effham}.\end{eqnarray}

The important parameter $U_0=g_0^2/\Delta_a$ here is the optical lattice depth per photon~\cite{Domokos03} and also gives the refractive index of one atom at an antinode. The term containing $\eta_{\textrm{eff}}=g_0h_0\zeta/\Delta_a$ represents an effective pump through atomic scattering into the mode. Along $x$ the cavity field forms an optical lattice potential with period $\lambda/2$  and depth $\hbar U_0 a^\dag a$. For the sake of generality we add an extra classical potential $V_{\textrm{cl}}$ as well.  

To derive a generalized Bose-Hubbard Hamiltonian we expand the Bloch states of a single atom inside the lattice using localized Wannier functions~\cite{kittel} and rewrite the field operators in Eq.~(\ref{ham2ndquant}) in these functions, keeping only the lowest vibrational state at each site (lowest band) $\Psi\left({\bf x}\right)=\sum_ib_i w\left( {\bf x}-{\bf x}_i\right)$ to get:
\begin{eqnarray}
\label{BHham} \nonumber H&=&\sum_{k,l}E_{k,l}b_k^\dag b_l + \left(\hbar U_0 a^\dag a + V_{\textrm{cl}}\right)\sum_{k,l}J_{k,l}b_k^\dag b_l\\\nonumber&+&\hbar\eta_{\textrm{eff}}\left( a+a^\dag\right)\sum_{k,l}\tilde{J}_{k,l}b_k^\dag b_l-i\hbar\eta\left( a-a^\dag\right)\\&+&\frac{U}{2}\sum_kb_k^\dag b_k\left( b_k^\dag b_k-1\right)-\hbar\Delta_c a^\dag a. 
\end{eqnarray}

The operators $b_k^\dag$ ($b_k$) correspond to the creation (annihilation) of an atom at site $k$ and the on-site interaction of two atoms is given by $ U=\frac{4\pi a_s\hbar^2}{m}\int\mathrm{d}^3 x\left\vert\left({\bf x}\right)\right\vert^4. $  As the nonlinear part of the nearest-neighbor interaction is typically two orders of magnitude smaller than the on-site interaction it is neglected as usually. In contrast to the classical field case~\cite{Jaksch} the appearance of the cavity field operator does not allow to reassemble all hopping terms to a single expression. To be still able to proceed analytically we assume a weak dependence of $w(x)$ on the cavity photon number. Although the opposite limit might even contain more interesting physics, we will concentrate on this limit to be able to proceed analytically. Explicitly the coupling matrix elements read:

\begin{subequations}
\begin{align}
 E_{k,l} = & \int\mathrm{d}^3 x\, w\left( {\bf x} - {\bf x}_k\right)\left( - \frac{\hbar^2}{2m}\nabla^2\right) w\left({\bf x} - {\bf x}_l\right),\\
J_{k,l} = & \int\mathrm{d}^3 x\,w\left({\bf x} - {\bf x}_k\right)\cos^2(kx)w\left({\bf x} - {\bf x}_l\right),\\
\tilde{J}_{k,l}  = & \int\mathrm{d}^3 x\,w\left({\bf x} - {\bf x}_k\right)\cos(kx)w\left({\bf x} - {\bf x}_l\right).
\end{align}
\end{subequations}

These matrix elements are symmetric, i.e., $E_{k,l}=E_{l,k},J_{k,l}=J_{l,k}$ and $\tilde{J}_{k,l}=\tilde{J}_{l,k}$ and the on-site elements $E_{k,k}$ and $J_{k,k}$ are independent of $k$. As the next-nearest neighbor terms are typically two orders of magnitude smaller than the nearest-neighbor amplitudes they are omitted too.  Note that in the case of transverse pumping two adjacent wells acquire different depths since the $\cos$ in~(\ref{effham}) changes sign periodically, which implies $\tilde{J}_{k,k}=-\tilde{J}_{k+1,k+1}$. The Hamiltonian~(\ref{BHham}) now reads:
    
\begin{eqnarray}\label{BHham2} \nonumber H&=& E_0 \hat{N}+E\hat{B}+\left( \hbar U_0 a^\dag a+V_{\textrm{cl}}\right)\left( J_0\hat{N}+ J\hat{B}\right) \\\nonumber&+&\hbar\eta_{\textrm{eff}} \left( a+a^\dag\right)\tilde{J}_{0}\sum_{k}(-1)^{k+1}\hat{n}_k-\hbar\Delta_c a^\dag a
\\&-&i\hbar\eta\left( a-a^\dag\right)+\frac{U}{2}\sum_k\hat{n}_k\left(\hat{n}_k-1\right), \end{eqnarray}

where we introduced $\hat{N}=\sum_k\hat{n_k}=\sum_k b_k^\dag b_k$ (number operator) and $\hat{B}=\sum_{k}\left( b_{k+1}^\dag b_k+h.c.\right) $ (jump operator). $E_0, J_0$ and $\tilde{J}_0$ are on-site matrix elements, whereas $E$ and $J$ are the site-to-site hopping elements. The Hamiltonian~(\ref{BHham2}) is a central result of this work and gives the starting point to discuss the physics below.

Lets first look at the light field dynamics and write down the corresponding Heisenberg equation:

\begin{eqnarray}\label{heisenberg_equ} \dot{a}&=& \left[i\left(\Delta_c-U_0\left( J_0\hat{N}+J\hat{B}\right)\right) -\kappa\right] a+\eta\nonumber\\&-&i\eta_{\textrm{eff}} \tilde{J}_{0}\sum_k(-1)^{k+1}\hat{n}_k.\end{eqnarray} 
We see that the quantum state of the field depends not only on the number of atoms $\hat{N}$ but also on coherences via $\hat{B}$. Particularly interesting effects can also be expected from the last term describing transverse pumping as the corresponding operator has an alternating sign for neighboring wells. Hence it vanishes exactly for a Mott insulator state, while it gives a nonzero contribution for a superfluid state. Nevertheless we will concentrate here on the more simple setup pumping acts via a cavity mirror and set $\eta_{\textrm{eff}}=0$ below. 


{\sl Bad cavity limit:} The common interaction of all atoms with the same field implies a complicated dynamics. Luckily in typical setups the field damping rate $\kappa$ is the fastest time scale in the system. This allows to eliminate the cavity degrees of freedom by formally solving equation~(\ref{heisenberg_equ}) for 
$\label{equ:a_s} a =\eta/[ \kappa-i(\Delta_c-U_0( J_0\hat{N}+J\hat{B})) ]$ and inserting this back to~(\ref{BHham}). As $\hat{N}$ commutes with $\hat{B}$ this gives no ordering problem. For a fixed atom number $N=\langle\hat{N}\rangle$ we then expand $a$ to second order in the small tunneling matrix element $J$:

\begin{equation} a \approx \frac{\eta}{\kappa -i\Delta^\prime_c}\left[ 1 - \frac{iU_0J}{\kappa -i\Delta^\prime_c}\hat{B} -\frac{U_0^2J^2}{(\kappa -i\Delta^\prime_c)^2}\hat{B}^2\right] \label{exp},\end{equation}

where $\Delta^\prime_c = \Delta_c - U_0J_0N $ is a rescaled detuning, so that we have:
\begin{eqnarray}
\label{BHham3}  H&=& \left[ E +  J\left( V_{\textrm{cl}} - \hbar U_0\eta^2\frac{\kappa^2 - {\Delta_c^\prime}^2}{\left(\kappa^2 + {\Delta_c^\prime}^2\right)^2}\right)\right]\hat{B}\\&-&3\hbar U_0^2\eta^2\Delta_c^\prime \frac{3\kappa^2 - {\Delta_c^\prime}^2}{\left(\kappa^2 + {\Delta_c^\prime}^2\right)^3} J^2\hat{B}^2\nonumber + \frac{U}{2}\sum_k\hat{n}_k\left(\hat{n}_k-1\right). \end{eqnarray}

\begin{figure}[htp]
\begin{center}
    \includegraphics[width=0.49\hsize]{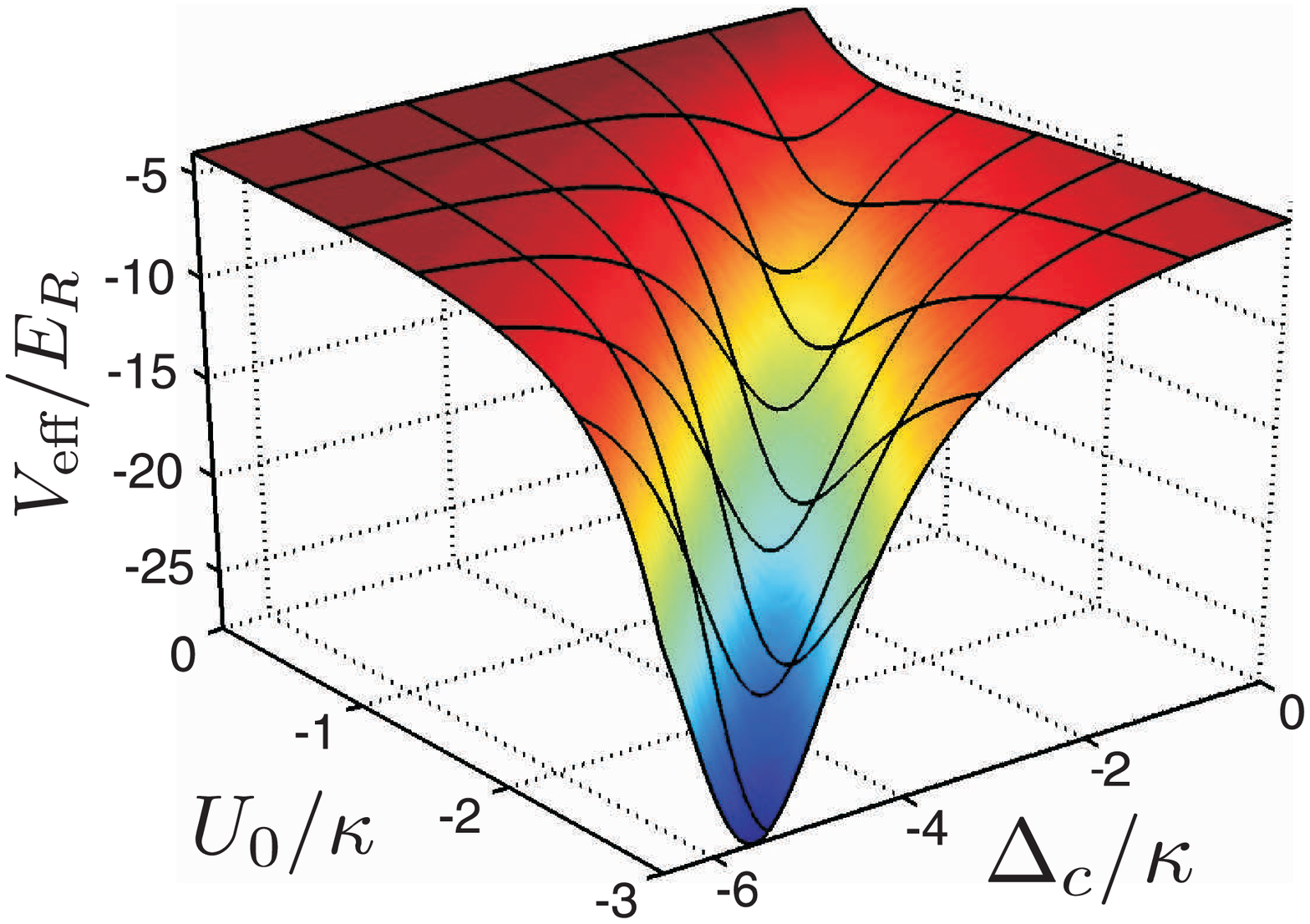}
     \includegraphics[width=0.49\hsize]{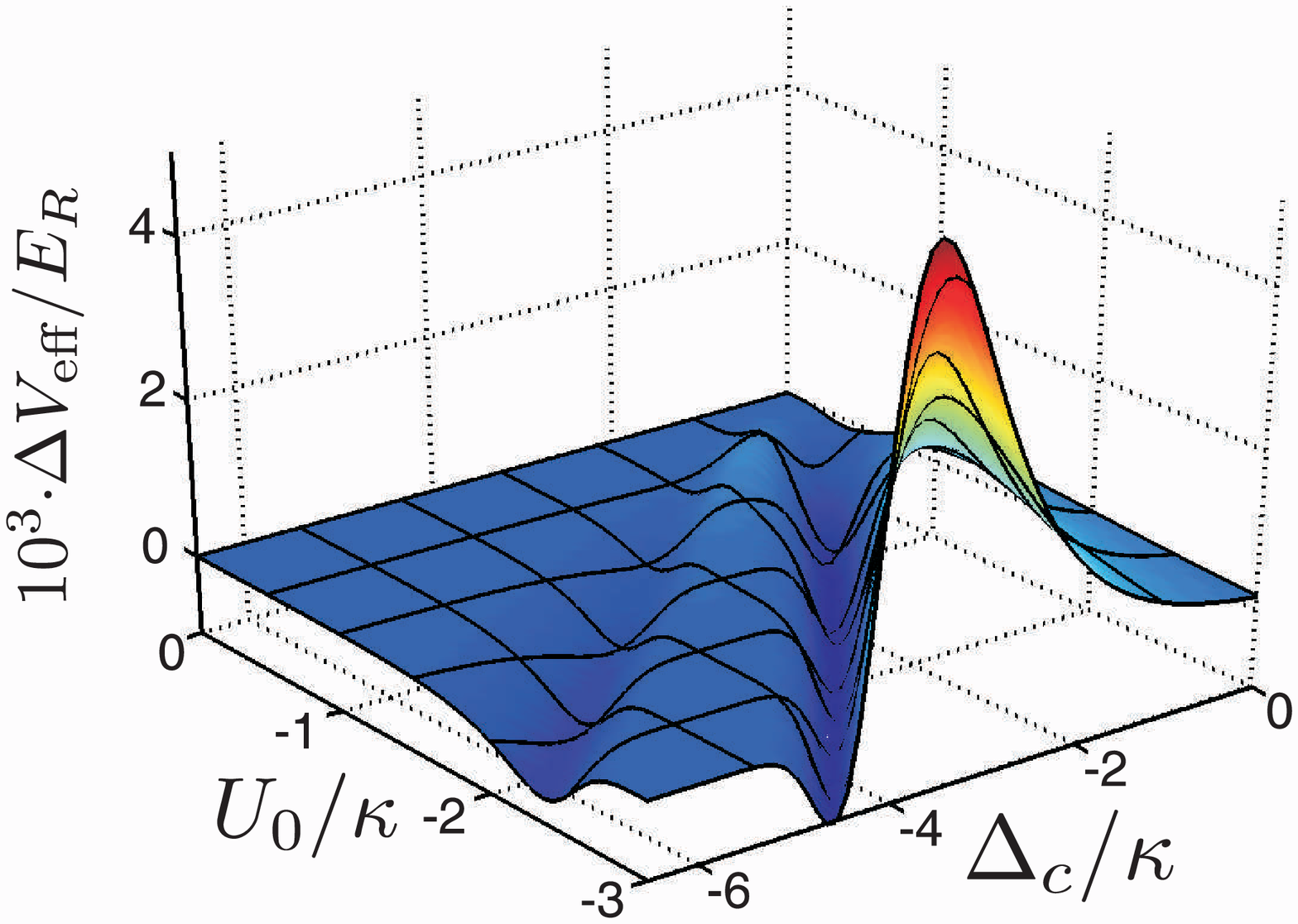}
  \caption{\label{fig:pot_depth}(color online). (a) Numerically found average potential depth $V_{\textrm{eff}} = V_{\textrm{cl}}+ \hbar U_0\langle a^\dag a\rangle$ of the ground state of~(\ref{BHham2}) for $N=2$ atoms in two wells. (b) Relative error in this potential using expansion~(\ref{BHham3}). The parameters are $\eta=2\kappa,\,V_{\textrm{cl}}=-4E_R$ and the scattering length is $a_s = 0.1E_R$.}
    \nobreak\medskip
  \end{center}
\end{figure}

Obviously $H$ now contains cavity induced rescaling of the tunnel coupling proportional to $\hat{B}$ as well as nonlocal correlated two atom hopping terms proportional to $\hat{B}^2$ mediating long range interactions. In Fig.~\ref{fig:pot_depth} we show the excellent agreement of the field calculated from the expansion~(\ref{exp}) compared to a full numerical inversion within a large parameter range as used below. Note that the matrix elements still weakly depend on the depth of the optical potential via $\langle \hat{B} \rangle$ but as $J$ is small they can be approximated neglecting the $\hat{B}$-term and setting $J_0 = 1$ in the field expectation value $\alpha^0 = \eta/[\kappa -i(\Delta_c-NU_0)]$.  

\begin{figure}[tp]
\begin{center}
    \scalebox{0.5}[0.5]{\includegraphics{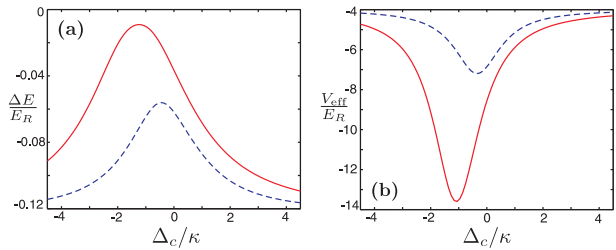}}
 \caption{\label{fig:delta_E}(color online). (a) Energy difference $\Delta E$ as function of $\Delta_c$ for a single atom in two wells for $U_0=-1.2\kappa$ (solid line) and $U_0=-0.4\kappa$ (dashed line). The associated lattice depth is shown in (b). The other parameters are $\eta=2\kappa$ and $V_{\textrm{cl}}=-4E_R$.} 
    \nobreak\medskip
  \end{center}
\end{figure}

Lets now discuss some key physics. In the simplest case of a single particle in two wells the symmetric and antisymmetric superpositions of the atom in either site are eigenstates with an energy difference

\begin{equation}\label{energy_diff} \Delta E = 2\left[ E + J\left( V_{\textrm{cl}} - \hbar U_0 \eta^2 \frac{\kappa^2 - {\Delta
_c^\prime}^2}{\left(\kappa^2 + {\Delta_c^\prime}^2\right)^2}\right)\right], \end{equation}

strongly depending on the cavity parameters (see Fig.~\ref{fig:delta_E}a). Hence detuning gives a handle to control the tunnel coupling and atom confinement (Fig.~\ref{fig:delta_E}b). Note that the symmetric and antisymmetric eigenstate are associated with different field amplitudes (lattice depths).

\begin{figure}[tp]
\begin{center}
       \scalebox{0.47}[0.5]{\includegraphics{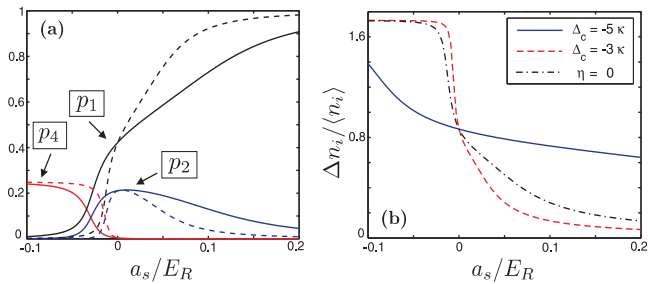}}   
 \caption{\label{fig:4atom}(color online). (a) Occupation probabilities $p_i$ for $i$ atoms in one well as function of scattering length for the ground state of 4 atoms in 4 wells. The parameters are $U_0=-\kappa$ and $\Delta_c = -3.75\kappa$ and $V_{\textrm{eff}} = \hbar U_0\langle a^\dag a\rangle = -4E_R$.  For comparison the dashed lines correspond to an equivalent classical potential, i.e., $\langle a^\dag a\rangle = 0$ and $V_{\textrm{cl}} = -4E_R$.
   (b) Fluctuations of atom number in one well as a function of $a_s$ for different cavity detunings $\Delta_c=-5\kappa$ (solid line), $\Delta_c = -3\kappa$ (dashed line) and a classical field (dashed-dotted line).} 
 \end{center}
\end{figure}

Adding more atoms the interaction term comes to play and the ground state of the system is a superposition of different atomic configurations. Here the cavity parameters influence the position and shape of the well known Mott-insulator superfluid transition~\cite{Fisher,Jaksch,Zwerger}. An important quantum feature appears for fields where the uncertainty in the photon number is not neglible. For an average photon number $\bar{n}$ generating a potential depth close to the phase transition point the photon numbers $\bar{n}\pm 1$ are than associated to different atomic phases, so that the ground state contains atomic states of different phases correlated with the corresponding photon number. Even for a system being dominantly in the insulator phase, photon number fluctuations then allow the atoms to jump. This is shown in Fig.~\ref{fig:4atom}a for 4 particles in 4 wells. Here we compare the site occupation probabilities as a function of scattering length for a purely classical and a quantum potential where the photon number uncertainty allows hopping even in the insulator regime. This behaviour can be enhanced or reduced through cavity mediated interaction as shown in  Fig.~\ref{fig:4atom}b, where we plot the atom number fluctuations in one well as a function of $a_s$ for different atom-cavity detunings and compare it to the classical field case. Clearly the atom number fluctuations are enhanced on one side of the cavity resonance and suppressed on the other. 

\begin{figure}[tp]
\begin{center}
         \scalebox{0.5}[0.5]{\includegraphics{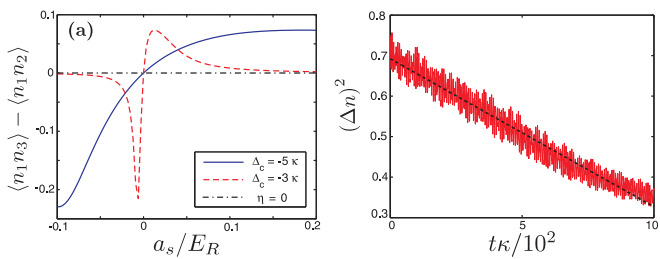}} 
     \caption{\label{fig:4atom_corr} (color online). (a) Difference of the density-density correlation functions: $\langle n_1 n_3\rangle - \langle n_1 n_2\rangle$. Parameters are as in Fig.~\ref{fig:4atom}b. (b) Dynamical evolution of atom number fluctuations in one well starting from an interaction free ground state and a sudden turn on of onsite interaction. A linear fit is depicted by the dashed line. Here we chose $U_0 = -\kappa,\, \Delta_c = -4.2\kappa,\,V_{\textrm{cl}}=0,\,a_s=3E_R$ and $\eta$ such that $V_{\rm eff}=-4E_R$.}
    \nobreak\medskip
  \end{center}
\end{figure}

The appearance of long range 4 particle interactions mediated by the $\hat{B}^2$-term in~(\ref{BHham3}) can be seen by comparing the density-density correlation functions $\langle n_i n_j\rangle$~\cite{denscorr} for nearest and next nearest neighbor sites. Depending on cavity parameters as shown in  Fig.~\ref{fig:4atom_corr}a each of the two correlations can be enhanced or reduced with respect to the classical potential case.


{\sl Coupled atom-field dynamics in the semiclassical limit:}\label{dynamics}
As final point we turn to the classical field limit of the coupled Hamiltonian~(\ref{BHham2}) for fixed $N$ and large photon number. Here the field is approximately in a coherent state $|\alpha\rangle$ and the system is assumed to evolve as a product state $|\Phi\rangle=|\psi\rangle\otimes|\alpha(t)\rangle$. The Heisenberg equation~(\ref{heisenberg_equ}) for the field then reduces to a classical equation for $\alpha$ containing expectation values of atomic operators: 

\begin{eqnarray}
\label{heisenberg_equ_cl} \dot{\alpha}(t) &=& \left[ i\left(\Delta_c - U_0\langle\psi| J_0\hat{N} + J\hat{B} |\psi\rangle\right) -\kappa\right] \alpha(t) + \eta, \nonumber\\
\nonumber i\hbar\frac{\mathrm{d}}{\mathrm{d} t}|\psi\rangle &=& \left[ E + J\left( V_{\textrm{cl}} + \hbar U_0\left\vert\alpha(t)\right\vert^2\right)\right]\hat{B}|\psi\rangle\\&+& \frac{U}{2}\sum_k\hat{n}_k\left(\hat{n}_k-1\right)|\psi\rangle.
 \end{eqnarray}

$\alpha(t)$ is then inserted back into the atomic Hamiltonian like a classical time dependent potential $V_{\textrm{cl}}$~\cite{HorakBEC,masch04}. Similar to the case of a time dependent Gross-Pitaevskii equation~\cite{HorakBEC} the corresponding Schr\"{o}dinger equation can be solved simultaneously, where the matrix elements $E,J$ have to be recalculated in each time step. 

Although the factorizing assumption is in general doubtful and one has to check the dynamical restriction to the lowest band, this procedure gives a first insight in the dynamical behavior of the model. As a generic example we show the time evolution of the uncertainty of the site occupation number starting with a 'superfluid' state at $t=0$, when the onsite interaction is turned on. Fig.~\ref{fig:4atom_corr}b shows that in contrast to a fixed external potential the dynamic cavity field leads to a damping of the fluctuations approaching a Mott insulator state.  

In summary we have shown that a dynamical quantum optical potential for ultracold atoms invokes a wealth of new physics. The effects are pronounced in the limit of strong coupling and small photon numbers but long range interactions persist even in the bad cavity limit within a classical field approximation. The considered systems are in range of current experimental capabilities and should allow to control and study new quantum phases.    

{\sl Acknowledgments:}
We thank G. Moriggi, M. Lewenstein and P. Domokos for helpful discussions. Funded by the Austrian Science Fund FWF - P17708.

\newpage

\begin{thebibliography}{99}
\bibitem{lasercool} See e.g. {\it Laser Manipulation of Atoms and Ions}, ed. by E. Arimondo and W. D. Phillips,Varenna Summer School, 1991 (North-Holland, Amsterdam 1992).
\bibitem{Fisher}M.\ P.\ A.\ Fisher {\it et al.}, Phys.\ Rev.\ B {\bf 40}, 546 (1989).
\bibitem{Jaksch} D.\ Jaksch {\it et al.,} Phys.\ Rev.\ Lett. {\bf 81}, 3108 (1998); D.\ Jaksch and P.\ Zoller, Ann.\ Phys.\ {\bf 315}, 52 (2005).
\bibitem{Zwerger}W.\ Zwerger, J.\ Opt.\ B  {\bf 5}, S9 (2003).
\bibitem{Expmott} M.\ Greiner {\it et al.}, Nature {\bf 415}, 39 (2002).
\bibitem{Domokos03}P.\ Domokos and H.\ Ritsch, J.\ Opt.\ Soc.\ Am.\ B {\bf 20}, 1098 (2003).
\bibitem{Griessner}A.\ Griessner, D.\ Jaksch, and P.\ Zoller, J.\ Phys.\ B {\bf 37}, 1419 (2004). 
\bibitem{BEC_Cav}B.\ Nagorny, T.\ Els\"{a}sser, and A.\ Hemmerich, Phys.\ Rev.\ Lett.\ {\bf 91}, 153003 (2003); J.\ A.\ Sauer {\it et al.}, Phys.\ Rev.\ A {\bf 69}, 051804 (2004).
\bibitem{masch04} C.\ Maschler and H.\ Ritsch, Opt. Comm. {\bf 243}, 145 (2004).
\bibitem{kittel} C.\ Kittel, {\it Quantum Theory of Solids,} John Wiley \& Sons, (New York 1963).
\bibitem{denscorr}T.\ D.\ K\"{u}hner, S.\ R.\ White, and H.\ Monien, Phys.\ Rev.\ B {\bf 61}, 12474 (2000).
\bibitem{HorakBEC} P.\ Horak and H.\ Ritsch, Phys.\ Rev.\ A {\bf 63}, 23603 (2001) 

\end{thebibliography}
\end{document}